\pdfoutput=1
\documentclass[10pt]{article} 

\usepackage{amsmath} 
\usepackage{amssymb} 
\usepackage{graphicx} 
\usepackage{cite} 
\usepackage{color} 
\usepackage[normalem]{ulem}

\usepackage[labelfont=bf,labelsep=period,justification=raggedright]{c
aption} 

\makeatletter
\renewcommand{\@biblabel}[1]{\quad#1.}
\makeatother

\date{ }
\begin{document} 
\begin{flushleft}
{\Large 
\textbf{The effects of spatially heterogeneous prey distributions on detection 
patterns in foraging seabirds}
}
\\
Octavio Miramontes$^{1,2,3,\ast}$, 
Denis Boyer$^{1,2}$, 
Frederic Bartumeus$^{4}$
\\
\bf{1} Departamento de Sistemas Complejos, Instituto de F\'isica, Universidad Nacional Aut\'onoma de M\'exico, 04510 D.F., M\'exico
\\
\bf{2} C3, Centro de Ciencias de la Complejidad, 
Universidad Nacional Aut\'onoma de M\'exico, 04510 D.F., M\'exico
\\
\bf{3} Departamento de F\'isica, UFPR, Curitiba, Brasil
\\
\bf{4} Centre d'Estudis Avan\c{c}ats de Blanes (CEAB-CSIC), 17300 Blanes, Girona, Espa\~na
\\
$\ast$ E-mail: octavio@fisica.unam.mx
\end{flushleft}

\section*{Abstract} 

Many  attempts  to  relate   animal  foraging  patterns  to  landscape
heterogeneity  are  focused on  the  analysis  of foragers  movements.
Resource  detection patterns  in  space  and time  
are  not  commonly
studied, yet they are tightly  coupled to landscape properties and add
relevant  information  on  foraging  behavior.   By  exploring  simple
foraging  models  in  unpredictable  environments  we  show  that  the
distribution of intervals between detected prey (detection statistics)
is mostly  determined by the spatial  structure of the  prey field and
essentially distinct from predator displacement statistics.
 Detections are  expected to be Poissonian in uniform
random   environments  for   markedly  different   foraging  movements
(\emph{e.g.}  L\'evy and ballistic).  This prediction is supported
  by data on  the time intervals between diving  events on short-range
  foraging  seabirds  such  as   the  thick-billed  murre  ({\it  Uria
    lomvia}). However, Poissonian detection statistics is not observed
  in  long-range  seabirds  such  as  the  wandering  albatross  ({\it
    Diomedea exulans})  due to the  fractal nature of the  prey field,
  covering  a wide  range of  spatial  scales.
For this  scenario, models  of fractal  prey  fields
induce  non-Poissonian patterns  of detection  in  good agreement
  with two  albatross data sets. We  find   that  the  specific  shape  
  of  the distribution  of  time  intervals between
prey detection  is mainly driven  by meso and  submeso-scale landscape
structures and  depends little on  the forager strategy  or behavioral
responses.

\section*{Introduction} 

A number seabird species search and catch prey in ranges from hundreds
 to   thousands   of    kilometers   away   from   their   nesting
sites \cite{prince1984activity,cairns1987activity,irons1998foraging,
gremillet2004offshore, berrow2000foraging, gonzalez2000foraging,
wood2000quantifying, crawford2008implications}.  The changing nature
of marine environments makes seabird prey distributions highly dynamic
and unpredictable  over large spatial scales,  ultimately impacting on
seabirds     capture     efficiency    \cite{weimerskirch2005foraging,
weimerskirch2005prey}.   In this  scenario, seabird  populations are
under  constant  survival pressure, a situation worsened by  climate
changes,  that  significantly   perturb  prey  availability  and 
the     ecology      of      predator-prey     systems
\cite{sydeman2001climate,weimerskirch2012changes}.   
A well known example is the impact  of 
El Ni\~no-ENSO oscillations in  the Pacific Ocean
\cite{mysak1986nino}  on  sardine  population fluctuations  off  South
Africa coast \cite{crawford2008implications}.  Studies of
how seabirds detect  and catch prey in the open ocean are also very 
important to assess   the   health   of   fish   stocks   
\cite{cairns1992bridging,monaghan1996relevance,
frederiksen2007seabirds,piatt2007introduction,einoder2009review},
particularly  for  declining  species  that are  commercially
valuable  \cite{pauly2003future,myers2003rapid}. The  availability of
telemetry       and        satellite       tracking       technologies
\cite{jouventin1990satellite,prince1992satellite,weimerskirch1993foraging,
viswanathan1996levy,phillips2004accuracy}
accounts for recent  progress in the understanding of  habitat use
and      foraging       behavior      of      long-range      oceanic
birds \cite{weimerskirch2000oceanic,louzao2011exploiting}.   Yet, this new empirical
knowledge has been seldom followed  up by  theoretical studies
providing general and more  formal rationale for the observed foraging
patterns.  Motivated by this,  and inspired by the long-range foraging
patterns  of albatrosses,  here we  explore how   landscape-properties    
({\it i.e.} large-scale  prey spatial distributions) affect prey detection 
patterns in seabirds.

Foraging              models             (see,             \emph{e.g.}
\cite{viswanathan1999optimizing,benhamou2007many,plank2008optimal})
often  examine  the  average  distance  (or  time)  travelled  between
successive prey  detections, a key quantity that is inversely  
proportional to the foraging efficiency. Much  less attention has been 
paid  to the entire distribution of distances/times between  detected 
prey (but see \cite{plank2008optimal}), herein referred to as detection  statistics. This latter quantity  has  been  sometimes directly measured, in 
particular        for wandering  
albatrosses (\emph{Diomedea exulans})
\cite{weimerskirch1997activity,weimerskirch2005prey}. It is worth noting that
detection
patterns  in  unpredictable environments  are  -\emph{a priori}-  not closely
related to displacement  patterns. Displacements,  \emph{i.e.}, a
set of  positions defining a  trajectory, reflect internal  states and
complex    behavioral    responses    to    resource    distributions
\cite{hassell1974aggregation,benhamou1989animals,with1999movement,fritz2003scale,nathan2008movement,sims2008scaling}.    
Detections,  in   turn,  are
localized events  resulting from the explicit  or physical interaction
of   the   forager   with    the   prey   field 
 and/or targeted landscape features.

For  the past decade,  a wide  debate has  focused on  animal movement
models with power-law move length distributions (L\'evy walks) and on
their possible interpretation as optimal search strategies of randomly
distributed  prey  
\cite{shlesinger1986levy,viswanathan1999optimizing,bartumeus2009behavioral,
reynolds2009levy,de2011levy}. The  movement patterns of
many foragers, for  instance, marine predators \cite{sims2008scaling},
plankton       \cite{bartumeus2003helical},       spider       monkeys
\cite{ramos2004levy}  or   jackals  \cite{atkinson2002scale}
display a wide range of spatial scales that cannot be accounted for by
Poisson  statistics. Wandering  albatrosses were actually one  of the first  
biological examples where evidence  for L\'evy displacements was reported
\cite{viswanathan1996levy,viswanathan1999optimizing}. Flaws found
later     in    the     analysis  questioned    these findings
and data of  higher  resolution
were neither fitted by a L\'evy law nor a Poisson law,
but by  a truncated modified power-law function \cite{edwards2007revisiting}. 
This set of studies has attempted to draw conclusions on the search 
strategies of albatrosses  
not from direct position tracking, but based on flight duration data, which 
were assumed  to be indicative of  detection times between prey 
\cite{viswanathan1996levy,viswanathan1999optimizing,edwards2007revisiting}. 
Here we provide further evidence showing that these data  are actually 
related to detections, but also show that they do not carry information 
on movements and, therefore, on the nature of the search patterns leading
to these detections.

We  more generally  examine  the  effects of  the  prey field  spatial
structure  and  of foraging  rules  on  detection  patterns.  For  prey
uniformly  distributed  in  space,  detection patterns  are  trivially
exponential if  displacements are  ballistic or self-avoiding  but the
outcome is less  obvious for other types  of movement.  We find
that L\'evy movement models \cite{viswanathan1999optimizing} also lead
to  exponential prey  detection patterns  in  Poissonian environments, which
illustrates the markedly different nature of detection
and movement statistics. These predictions can explain the diving patterns 
of short-range foraging seabirds, such as the thick-billed murre 
\cite{hamish2009underwater}, whose dives are exponentially distributed 
on time.

Prey in the ocean are not uniformly distributed at large scales, though 
\cite{rodhouse1996cephalopods,freon2005sustainable,garrison2002spatial,
tsuda1995fractal}. 
Detection patterns in complex media have been little studied 
and mostly in  non-biological  contexts \cite{isliker2003random}. 
We show that the non-Poissonian albatross data 
of \cite{weimerskirch2005prey}
and \cite{edwards2007revisiting} can  be explained  by models of a
forager flying over a fractal  prey landscape with parameter values
consistent with observed resource distributions in the ocean.
We  use   two  models  generating  fractal
landscapes of  different nature and relate the fluctuations
in the predator  detection  times (or distances) to the
prey density  heterogeneities. As in the uniform case, detection  patterns 
in a given environment are  found robust  with  respect to  a variety  of
foraging  rules, where  the predator  may  or may  not switch  between
different behaviors depending on prey detection.

\section*{Foraging seabirds: movement vs. detection statistics} 

In this study we re-analyze data from thick-billed murres
and wandering albatrosses, two seabird species with
markedly different behaviors. Thick-billed murres forage over small spatial
scales in short foraging trips (representing less than 1h of flight in
total) within a few kilometers of their colony \cite{hamish2009underwater}.
They feed on benthic or pelagic
fish in zones where prey occur in patch and are relatively predictable.
These animals show a high degree of site fidelity. 
Murres perform above-water and underwater searching, although the latter 
has a much shorter mean duration \cite{hamish2009underwater}.
In ref. \cite{hamish2009underwater}, flight durations ($t$) of thick-billed 
murres between consecutive dives were measured with time-depth-temperature 
recorders.

On the other hand,
telemetry  data  reveal  that   some  albatrosses  species,
especially  wandering   albatrosses,  perform  exploratory   trips  of
thousands  of kilometers  involving commuting  and looping  typical of
central-place                                                  foraging
\cite{weimerskirch2005prey,weimerskirch2007does}.   This  large  scale
behavior  is interspersed  with hierarchically  nested area-restricted
search induced  by the recognition of  water masses such  as the shelf
edge, seamounts  or frontal  zones.  Prey are  likely to  be scattered
within  these  mesoscale  physical  structures that  represent  higher
profitability areas  that need to be  prospected, involving successive
landings   and   take-offs  \cite{weimerskirch2007does}.    Heart-rate
recorder  signals  in wandering  albatrosses  show  that landings  and
take-offs represent  a high energy expenditure for  these large birds,
who practically consume as much energy flying with a favorable tail or
side  wind  as when  sitting  on  the water  or  resting  on the  nest
\cite{weimerskirch2000fast}.  From an optimality  standpoint landings
should be considered informed behavioral responses, mostly associated
to  prey detection or  exclusive seascape  features, but  not strictly
related       to        successful       prey       captures.       In
\cite{weimerskirch2005prey,weimerskirch2007seabirds}  it  was observed
that  birds need  about two  landings on  average before  capturing prey
(measured from stomach temperature  sensors data).  In particular, two
capture   modes  have  been   identified  in   wandering  albatrosses:
``foraging  in flight'',  where  the  prey is  captured  within a  few
seconds after landing, and ``sit-and-wait'', where the bird is sitting
on  the   water  for   more  than  10   min  before  prey   is  caught
\cite{weimerskirch1997activity,weimerskirch2007does}. The sit-and-wait
strategy appears to  be a secondary tactic used  for prey clustered in
small patches, for which foraging in flight would require high turning
and    landing    rates,    or    for   prey    capture    at    night
\cite{weimerskirch2007does}. Albatrosses also land in water
to rest,  probably selecting  the resting areas  as well.  
Herein the
term ``prey detection'' will denote  the detection of prey, prey cues,
or   targeted  seascape   areas  (for   prospection,   potential  prey
captures, resting, etc.) that may induce landing or diving responses.

 One of  the wandering albatross data discussed  in the following
  (Bird   Island   data    \cite{edwards2007revisiting})  were
  obtained  with  wet-dry  sensors  measuring  flight
durations  ($t$)  between  successive  take-offs  and  landings.   The
 data was acquired in 2004 with a reading each
$\Delta t=$10s  \cite{edwards2007revisiting}. However this technology,
which is similar to that of the murre data mentioned above,
does not  give information  on trajectories and  the animals  were not
equipped with a high resolution GPS device.  The second wandering
  albatross   data   set  re-analyzed   here   (Crozet  Islands   data
  \cite{weimerskirch2005prey}) consists  in distances between captured
  prey  measured using stomach  temperature transmitters  and position
  tracking systems.

Let us now consider, as an illustrative example, the  search  model of 
\cite{viswanathan1999optimizing}. A forager with constant velocity $v$ 
chooses  randomly   oriented,  rectilinear  displacements  of
lengths  ($l$) drawn  from a  probability distribution  function (PDF)
$P_0(l)$. Prey is  immobile and randomly,
uniformly distributed  on a  plane in number  density $\rho$,  and the
forager can detect a prey only when it is at a shorter distance than a
perception radius $r$. A step is  stopped if a prey is detected on the
way or  completed otherwise.  

Viswanathan \emph {et al.} \cite{viswanathan1999optimizing} considered power-law distributions,
$P_0(l)=Cl^{-\mu}$ for $l>l_0$ and zero otherwise, where
$1\le \mu\le 3$. To test this  move length distribution  
for wandering
albatrosses, \cite{viswanathan1999optimizing} and
\cite{edwards2007revisiting} compared the  PDF of the 
flight durations $t$   obtained  from   the   wet-dry  sensors   to  
a power-law distribution. A similar comparison was performed with 
the flight
duration data of the thick-billed murres \cite{hamish2009underwater}.
In these studies, $t$
was thus assumed to be indicative of  a chosen move
length, $l$. But, as we have argued,  $t$  represents the time
elapsed between two detections, not a time spent traveling in straight 
line between two re-orientations. As $t$ and $l$ are different variables,
they \emph{a priori} obey different distributions. Therefore, comparing  
the model (or any other foraging model) with the
three data sets described above
requires to seek the PDF  of the distance flown between two successive
detected  prey  for that  model,  denoted as  $L$  here  (if the  bird
velocity is constant, then $L=vt$). Equivalently, $L$ is the sum of the 
step lengths
travelled between prey. One may use  the identity $PDF(L)=-dp(L)/dL$,
where $p(L)$ is  the probability that a path of length $L$ has
not found a prey yet (or the fraction of flights of length $\geq L$).

\section*{Prey detections in Poissonian landscapes}

\begin{figure} [!ht]
\begin{center}
\includegraphics[width=6in]{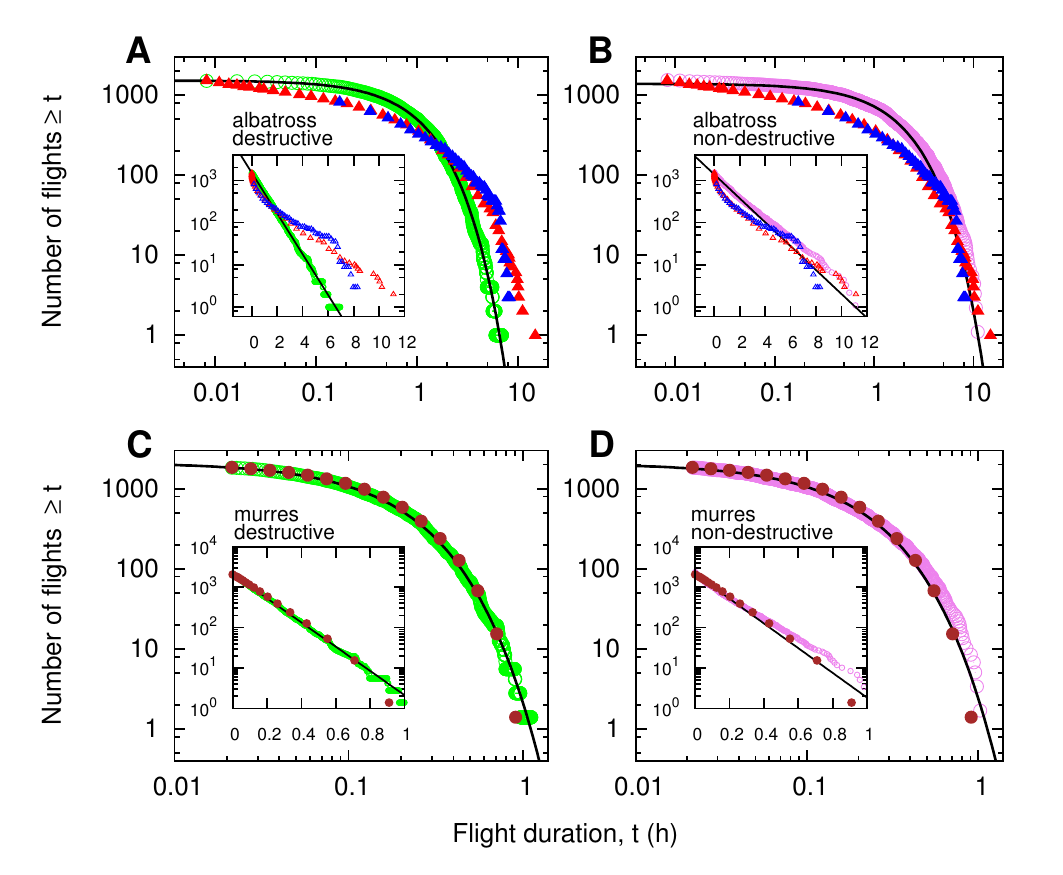}
\end{center}
\caption{Detections in random  uniform prey landscapes. (a) Albatross
  data  and  the  model  with destructive  scenario.   Green  circles:
  accumulated distribution $p(L)$ of flight lengths between successive
  detected prey of  the model forager \cite{viswanathan1999optimizing}
  with  perception radius  $r$=0.001 following  a L\'evy  process with
  $\mu$=1.5 (from  20 simulations of  75 captures each).  The foraging
  ground is represented  by a square of area  unity and contained 5000
  prey. Continuous line: exponential fit. Red triangles: $p(L)$ of the
  Bird        Island       albatross        takeoff/landing       data
  \cite{edwards2007revisiting}. Blue  triangles: $p(L)$ of  the Crozet
  Islands  albatross  prey  capture data  \cite{weimerskirch2005prey},
  converted into flight durations  assuming a constant flight velocity
  $v=16$m/s.   Inset: same  curves represented  in semi-log  to better
  emphasize the non exponential  nature of the observed albatross data
  versus the  exponential form of  the model forager  detections.  (b)
  Albatross  data  and model  with  non-destructive scenario.   Violet
  circles:  accumulated  distribution  $p(L)$  for the  model  forager
  performing  a   L\'evy  process  with   $\mu$=2.   Continuous  line:
  exponential fit. Prey  number: 3000; $r$=0.0003.  In a)  and b), the
  lengths in  the model with foraging arena of area unity  are converted in  hours ($t$)  by using
  $L=vt$  with  the  scaling  factor  $v$=0.12.   Inset:  same  curves
  represented  in semi-log  to  better emphasize  the non  exponential
  nature of the observed albatross data versus the exponential form of
  the  model  forager  detections.   (c)  Murre data  and  model  with
  destructive  scenario.    Green  circles:  accumulated  distribution
  $p(L)$  of flight  lengths between  prey of  the model  forager with
  perception  radius   $r$=0.001  following  a   L\'evy  process  with
  $\mu$=1.5  (from 20  simulations  of 75  captures each).  Continuous
  line:  exponential fit. Brown  dots are  the murre  flight durations
  from \cite{hamish2009underwater}.   Inset: same data  represented in
  semi-log in order to better emphasize the exponential nature of both
  the observed murres data and  the model forager.  (d) Murre data and
  model  with non-destructive  scenario.  Violet  circles: accumulated
  distribution $p(L)$  for a forager performing a  L\'evy process with
  $\mu$=2.  Continuous  line:  exponential  fit.  Prey  number:  3000;
  $r$=0.0003. Inset: same curves represented in semi-log. Similar
  close-to-exponential detections are obtained in all simulations
  with $1\leq \mu\leq 3$, in both destructive and non-destructive scenarii.}
\label{uniform_prey}
\end{figure}

We  illustrate below  that the  distance flown  between  to successive
detection events, i.e. $L$, is exponentially distributed in random and
uniform prey fields,  even if the choice distribution  $P_0(l)$ is not
an  exponential.   In  such  landscapes, if  detected  prey  disappear
(destructive scenario), any foraging  strategy producing paths that do
not revisit the same  location is optimal. Such non-oversampling paths
can be ballistic  (similar to a L\'evy process  with $\mu \approx 1$),
spirals,  self-avoiding walks,  etc...  Any  non-oversampling  path of
length $L$  has a  probability $p(L)=\exp(-L/\lambda)$ of  not finding
any prey,  with $\lambda=1/(2r\rho)$ a  characteristic distance, being
$r$ a  detection radius and $\rho$  the prey density.   If the forager
follows a random L\'evy search, its trajectory involves some degree of
oversampling.  We have obtained  $p(L)$ from numerical simulations for
this model.  In a destructive
scenario, in which prey are depleted and not revisited, $p(L)$ closely
follows  an exponential, not  only for  $\mu \approx  1$ but  also for
walks with $1< \mu \leq 2$ (Figure~\ref{uniform_prey}a and c). In the
non-destructive  scenario,  prey can  be  revisited.   If one  chooses
$\mu=2$ or any smaller  value, 
one also observes exponential detection
statistics  in a  very good  approximation (Figure~\ref{uniform_prey}b
and    d).     The    distribution    $-dp(L)/dL$   has    the    form
$\lambda_d^{-1}\exp(-L/\lambda_d)$,  a shape  which is  not  related to
that of $P_0(l)$.  These results illustrate that exponential tails for
prey detection statistics are an essential outcome of foraging models,
including those generated from L\'evy processes, when the landscape is
Poissonian.  However,  the precise value of  the characteristic length
travelled between prey, $\lambda_d$  (which is related to the foraging
efficiency),  generally depends  on  the scenario  and movement  rules
($P_0(l)$,  here).   As we  assume  that movement  
is  truncated  by detections, $\lambda_d$ is finite.

The  simple exponential form of  
$p(L)$ obtained for uniform prey fields describes well the murre data.
The maximum likelihood estimate (MLE) of $\lambda_d$ is $9.5$min and
a log-likelihood ratio test of goodness-of-fit (G-test) 
was performed from $10^4$ independent Monte Carlo samplings, giving 
$P=0.82$ ($n=2083$, $df=15$). In contrast, the exponential does not
describe none of the wandering albatross curves, see Figure~\ref{uniform_prey}
(G-test, Bird Island: $P<$ 0.0001,  $n$= 1507, $df=47$; Crozet Islands:
$P<0.0001$, $n=276$, $df=47$). In this figure, distances in the 
Crozet I. data were converted into flight durations assuming a constant flight
velocity $v=16$m/s \cite{alerstam1993flight,weimerskirch2007does}. The 
resulting curve lies very close to the Bird Island data.

\section*{Prey detection in large scale fractal landscapes}
 
In the case of Bird Island wandering albatrosses, Edwards {\it et al.} 
accurately fitted the flight duration distribution by a shifted
gamma function, which is asymptotically an exponential multiplied 
by an inverse power-law \cite{edwards2007revisiting}. Similarly, Weimerskirch
{\it et al.}  found   that   the
distribution of distances between  captured prey by Crozet Islands 
wandering albatrosses 
did not follow a simple exponential, but approximately
an inverse power-law \cite{weimerskirch2005prey} (see also 
\cite{sims2007minimizing}).

Such intermittent landing by albatrosses, often related to prey capture behaviour, 
can be explained by fractal
prey  landscapes. As a  matter of  fact, wandering  albatrosses forage
over much larger spatial scales than murres and
mainly feed on squid and  pelagic fish  \cite{weimerskirch2005prey}. This
prey 
display   several  levels   of  spatial   aggregation   and  schooling
\cite{rodhouse1996cephalopods,freon2005sustainable}  and  have  strong
spatial overlap with plankton \cite{garrison2002spatial}.  The large scale 
horizontal spatial
distributions   of  plankton  \cite{tsuda1995fractal,sims2008scaling},
passive      drifters      \cite{osborne1989fractal},      cephalopods
\cite{rodhouse1996cephalopods}         and         pelagic        fish
\cite{fauchald2000scale,bertrand2005levy,freon2005sustainable,makris2006fish}
are  known to  be self-similar  (with fractal  dimension  $D_F \approx
1.2-1.6  <2$) over  a wide  range of  scales, typically  from  a lower
characteristic scale $R_0$, of tens of meters, to an upper scale $R_m$,
of                             100$-$300                            km
\cite{tsuda1995fractal,fauchald2000scale,makris2006fish}.   At  scales
larger  than   $R_m$  the   prey  field  is   seen  as 
heterogeneous but space filling, that is, with $D_F \approx2$.
The 
mechanisms generating fractal horizontal distribution of marine species near
the ocean surface
are not well-known. 
Oceanic turbulence \cite{makris2006fish,tsuda1995fractal} 
and predator-prey interactions \cite{medvinsky2000fish} 
are two factors often invoked.

Based  on these  field observations,  we  consider  below  more realistic  
prey distribution models that generate fractals of different types.

\subsection*{(a) Truncated L\'evy Dust model (LD)}

It is  commonly accepted that  the assumption of  randomly distributed
prey in  spatial ecological models  is not entirely  appropriate since
there is a growing body of  evidence showing that prey are more likely
distributed in a patchy  and aggregated fashion.  This seems to
be especially true for distributions of prey in marine environments as
discussed                   above
\cite{fauchald2000scale,bertrand2005levy,freon2005sustainable,makris2006fish}.
L\'evy dusts in finite domains are a convenient method to generate stochastic 
fractal point patterns and they have been applied to model 
oceanic prey  fields  \cite{tsuda1995fractal,sims2008scaling}.
They have been less often used to model  the
movement of foragers profiting on these, though (but see \cite{bartumeus2008fractal}).
Our first  fractal foraging  model therefore employs  truncated L\'evy
dusts (LD)  to generate fractal prey  locations. 

\begin{figure}[!ht]
\centering 
\includegraphics[width=0.65\textwidth]{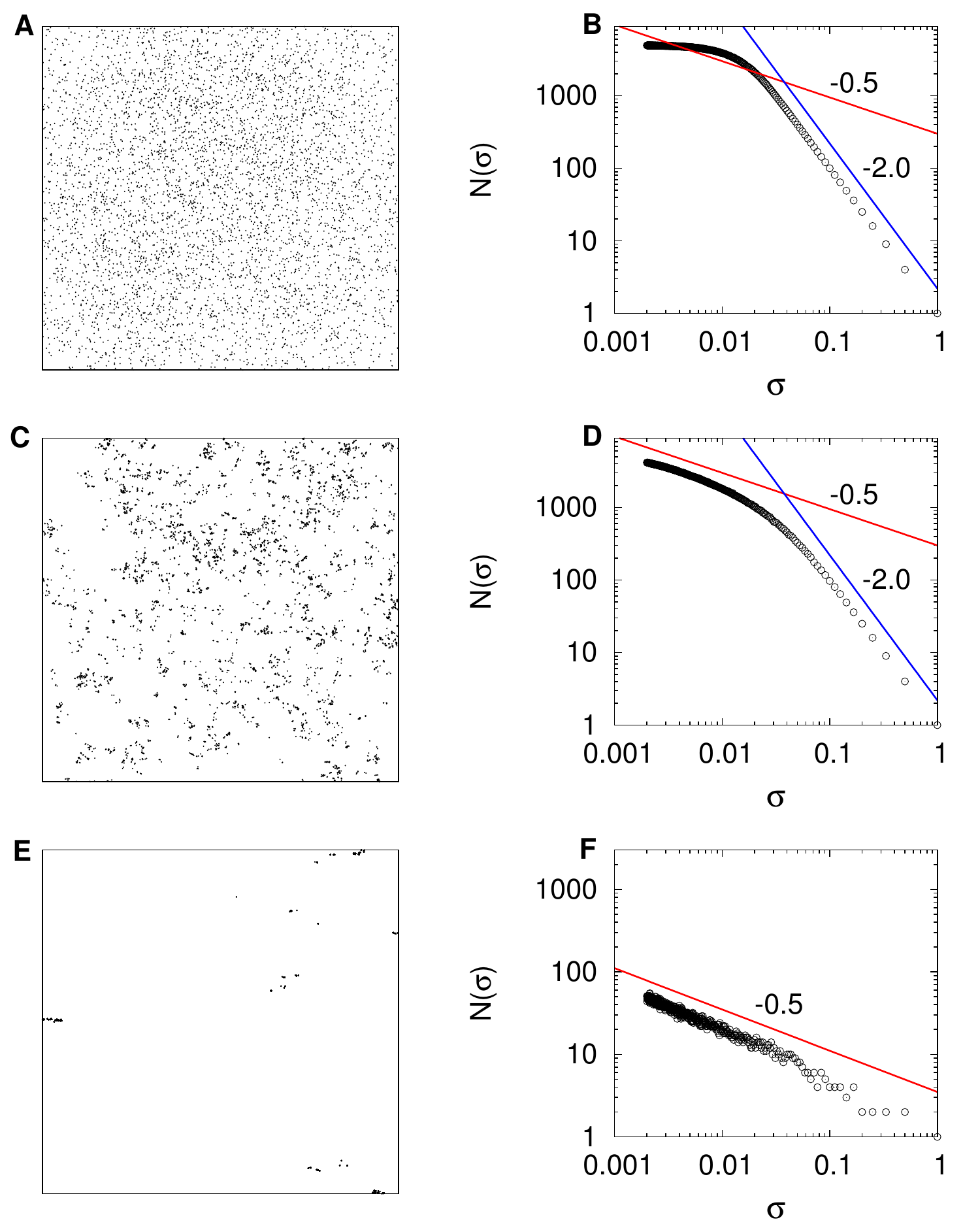}  
\caption{ 
Three different theoretical patterns of spatial prey distribution in a unit
box and their corresponding box-counting fractal dimension. $\sigma$ 
represent the size of the boxes and $N(\sigma)$ is the number of boxes 
of size $\sigma$ in the box-counting algorithm. In the three cases, 5000 
prey are distributed accordingly to a L\'evy dust with fractal dimension 
$D_F=0.5$ ($\beta=1.5$). (a) If the minimal distance between prey is large 
the L\'evy process producing the fractal pattern bounces many times 
on the walls and the overall process tends 
to be space-filling. In this particular case, the minimal distance 
between prey was $1/7$ and the process bounced around 2500 times which 
is equivalent to the superposition of 2500 separated fractals in the 
same domain. (b) As expected in this case, the fractal dimension measured 
by box-counting does not show a scaling region with 
exponent $D_F=0.5$ (red line) but approximates more the typical graph 
of a 2D random process with $D_F=2.0$ (blue line). (c) Pattern that 
corresponds to a prey distribution with a minimal distance of $1/700$ 
between prey, leading to less than 150 bounces
(~3\% of the total prey number). (d) In this case a scaling region with
$D_F=0.5$ is visible, followed by a two-dimensional behavior at larger 
length scales. (e) A very clumped and aggregated fractal pattern of 
prey is obtained when the minimal distance between 
prey is set to $7\times10^{-6}$. (f) In this case the fractal is nearly 
perfect with $D_F=0.5$.}
\label{levydust}
\end{figure}

LD are standard L\'evy  flights coming from the power-law distribution
$f(x)\sim x^{-\beta}$ with $1<\beta\leq3$ \cite{mandelbrot1982fractal}
and where only the turning points joining successive displacements $x$
are  considered as  prey locations.   This method generates
point patterns with  fractal dimension $D_F=\beta-1<2$ 
(Figure~\ref{levydust}). The  power-law  distribution  when   finite 
(contained  in  a square domain-box of unit length) is  truncated in 
the range $[\delta,1]$
where $\delta$ is interpreted  as the minimum distance  separating 
neighboring prey.   On  the other  hand,  the  maximum distance
separating consecutively located prey   is   the   domain-box   size,  
set   to   1 for convenience. Between both limits  (which define the 
self-similarity range of the fractal) the corresponding   truncated   
probability   distribution   function   is normalized to 1. 
The L\'evy dust  generator starts at the center of  a square domain of
unitary   area    and   accommodates   $N$    successive   prey   (see
Figure~\ref{levydust}).   When a new  prey position  is to  be located
outside of the domain, it is discarded and a new step is attempted (we
call  this a  \lq\lq border-bounce").   The fractal  nature of the 
pattern  may disappear if the  number of bounces is too high.  
In order to prevent
this, a tuning of $\delta$ is  applied to guarantee that the number of
bounces  is low, given a total number of prey. If the  distance $\delta$  
is large enough (but  always $\ll1$) and if  the total number of prey  
is also large, the  pattern approximates a Poisson distribution  because 
of  too much bouncing. If the value of $\delta$ is too small, the prey field
is limited to a very small region of the domain.  
An intermediate situation would produce a locally sparse fractal
covering  the whole  domain.   In  our simulations,  we  took values  of
$\delta$  such that  a bounce  occurs in  no  more than  1.5\% of  the
total number of prey.  It is also  necessary to keep
in mind that the value of $\delta$ depends on the value of the scaling
exponent  $\beta$ of  the  walker.  The  lower the  exponent
$\beta$ is,  the lower  the value of  $\delta$ has  to be in  order to
generate   an   undistorted    fractal   with   few   bounces   (see
Figure~\ref{levydust}). We  will discuss in the  following section the
detection  dynamics  of a  forager  moving  on  a fractal  prey  field
generated by this process (see Figure~\ref{ld_vs_fld}, left).

\subsection*{(b) Fractal Local Density (FLD) model}

We next propose an alternate and original model that builds stochastic fractals 
where,  in  contrast with  L\'evy  dusts  or Sierpinski-like   hierarchical   
structures   \cite{fauchald2000scale,isliker2003random}, the local prey 
density $\rho$ is well-defined.

Acoustic  devices allow to  measure the  density of  marine organisms,
either  locally  (e.g.,\cite{sims2008scaling})  or  over  hundreds  of
kilometers   squared   instantaneously  \cite{makris2006fish}.   Krill
density  has been  observed to  fluctuate widely from one location to another 
and to  follow a  power-law frequency  distribution,  of   the  form  
$PDF(\rho)\sim\rho^{-\alpha_\rho}$,  with $\alpha_\rho   \approx   1.7$
\cite{sims2008scaling}.  These large density variations also have a spatial 
structure that involves many length scales  across the 
landscape \cite{makris2006fish}. Therefore, to
characterize the prey field as a patchwork of regions with different 
densities, one must specify the sizes  of  these regions. 
These length scales ($R$ below) represent an other important 
ingredient of the model, as the local density alone is  not a  space  variable.
For albatrosses, fairly  localized high  productivity marine
areas occur  interspersed with vast oceanic areas  of low productivity
\cite{weimerskirch2005prey,weimerskirch2007seabirds}. We construct a model
that captures these properties. In the model,
high density regions  are numerous but small, and  represent overall 
a small fraction  of the  total area, corresponding  to the tail  of the
density PDF. On the contrary,  a significant area fraction is occupied
by a  few large regions  of very low  local density. The density  is a
continuous variable bounded by a minimal and a maximal value.

\begin{figure} [!ht]
\centering 
\includegraphics[width=6in]{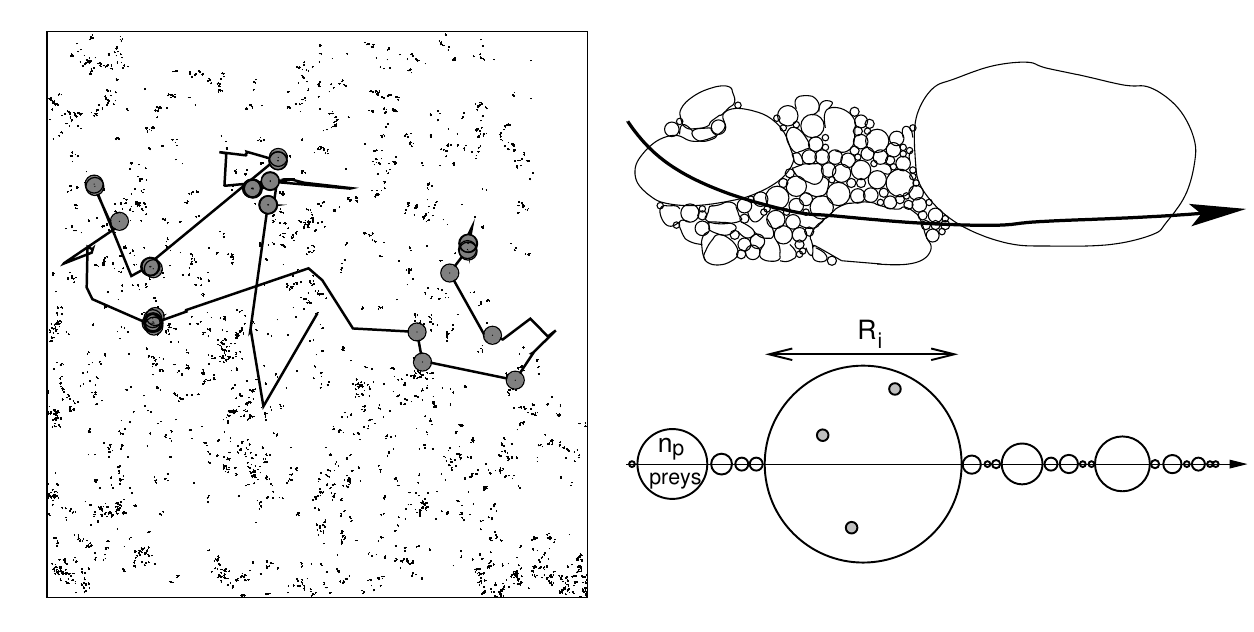}
\caption{
Left panel. Foraging arena composed of $N$=5000 prey generated with 
a LD of exponent $\beta$=1.5 (fractal dimension $D_F$=0.5). 
Solid line: trajectory of a ballistic forager ($\mu\simeq 1$) with 
detection radius $r$=0.001. The larger grey dots indicate detection 
events (destructive scenario).  Right panels: Fractal Local Density 
(FLD) model. Upper figure: The medium is composed of patches of heterogeneous
 sizes $R$, drawn from a PDF $\psi(R) \propto R^{-\nu}exp(-R/R_m)$. 
 Within a patch, $n_p(R)\propto R^{\epsilon}$ prey are randomly and 
 uniformly distributed. Lower figure: linear representation of the 
 forager/medium system, which is solved here.}
\label{ld_vs_fld} 
\end{figure}

The definition  of a patch tends  to be rather  inclusive. We define
here a  patch as a  region of space  of uniform prey density,  with no
limitation  on   its  size  and   density  
\cite{levin1974disturbance,kotliar1990multiple}. Consider a random assembly 
of non-overlapping,
roughly circular patches of varying  diameters $R$ that are drawn from
a    frequency    distribution   $\psi(R)$    (Figure~\ref{ld_vs_fld},
right). Inside a patch of size $R$, an average number of $n_p(R)$ prey
are distributed  randomly and uniformly. Therefore, the density in a 
patch is proportional to $n_p(R)/R^2$. To obtain a medium with
fractal properties up to a scale $R_m$, one first distributes $R$ according
to a truncated power-law distribution:
\begin{equation}\label{pdfR}
\psi(R)=cR^{-\nu}\exp(-R/R_m),\quad {\rm with}\ R\ge R_0
\end{equation}
where $\nu$ is  an exponent related to the fractal dimension, $R_m$  
the large cut-off length  of the
fractal  mentioned earlier,  and $c$  the normalization  constant. 
In addition, $R$ is always larger than some length $R_0$, which is the
minimum size  of a patch $(R_0\ll R_m)$ and can be taken  as the resolution
size. We next  assume an algebraic relationship between  the size of a
patch and the number of prey it contains:
\begin{equation}\label{eps}
n_p(R)=kR^{\epsilon},
\end{equation}
with  $k$ a constant  and $\epsilon$  an exponent  $\leq 2$. The case
$\epsilon  =2$ corresponds  to a uniform Poissonian  medium, where all regions 
have the same density. As  further
shown,  the  albatross  data  is  best fitted  by  landscapes  with
negative  values of  $\epsilon$:  large patches  have  fewer prey.  On
length  scales  $R_0  < R  \ll  R_m$,  the  patch distribution  (1)  is
scale-free, whereas practically  no patch has a size  much larger than
$R_m$. The box-counting method shows that  
for some parameters $\nu$ and $\epsilon$, the prey distribution of this model 
forms  a fractal set with dimension $D_F<2$ on  scales smaller than $R_m$ 
(see Supplementary Information). Restricting ourselves to the case  
$\epsilon<0$ of  interest  here, one  finds  that the fractal 
dimension is given by:
\begin{equation}\label{scaling}
D_F=\nu-\epsilon-1<2,\quad {\rm for}\ \nu<3+\epsilon,
\end{equation}
whereas $D_F=2$ for $\nu>3+\epsilon$. When $D_F<2$, the PDF of the local 
prey density is an inverse power-law (over a wide range of 
densities provided that $R_m/R_0$ is sufficiently large), with exponent 
given by:
\begin{equation}\label{scalingrho}
\alpha_{\rho}=\frac{5-\epsilon-\nu}{2-\epsilon}>0.
\end{equation}
In  this medium, we  consider the  case of  a ballistic  predator with
constant  velocity  $v$. Ballistic  motion  is  the simplest  movement
behavior  and  can  accurately  represent  albatross  relocations  at
certain             scales             \cite{weimerskirch1997activity,weimerskirch2007seabirds}.   If  we   assume  that   there   are  no
correlations between  the sizes  of neighboring patches,  the problem
can   be  simplified   to  that   of  a   forager  flying   through  a
one-dimensional   succession   of   patches   (Figure~\ref{ld_vs_fld},
right).  The process  is  easy  to
simulate numerically: during an  elementary time step $\Delta t$(=10s,
as in \cite{edwards2007revisiting}), the  forager located in a patch of size
$R_i$  travels  a  distance  $R_0=v\Delta  t$ and  has  therefore 
a Poissonian probability $\exp[-\tilde{r}(R_i/R_0)^{\epsilon-2}]$ of not 
finding any prey, with $\tilde{r}=2rv\Delta tk  R_0^{\epsilon-2}$ the 
dimensionless detection radius  and $R_i/R_0$  the dimensionless  patch size.  
The  process is iterated until  the end  of a patch  is reached,  
when a new  $R_i$ (and therefore a new prey density) is drawn from 
Equation~\ref{pdfR}.

After a  prey is detected, the  forager can either (i)  follow its way
(``non-responsive  search'') or (ii)  stay within  the same  patch for
$R_i/R _0$  other elementary  time steps (``responsive  search''). The
latter rule  mimics area restricted  search \cite{benhamou1989animals,plank2008optimal}, a  behavior that has been  observed in wandering
albatrosses \cite{weimerskirch1997activity,weimerskirch2007seabirds}.  
With  rule (ii),  the  forager tends  to
exploit more intensively higher  density regions, where detections are
more probable.

\subsection*{Results}

\begin{figure} [!ht]
\centering
\includegraphics[width=6.5in]{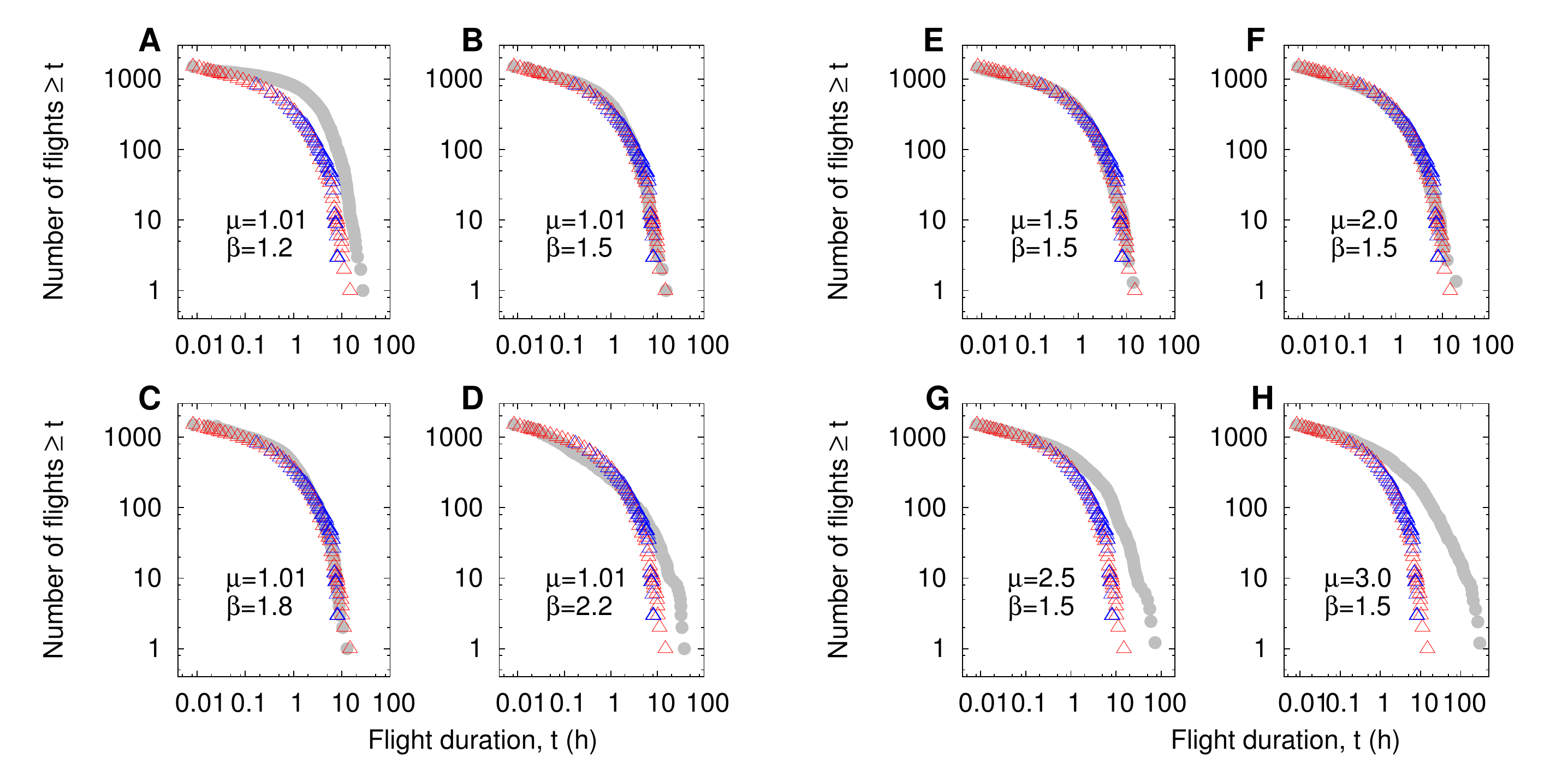}
\caption{
Detections in LD media.
(a)-(d): Accumulated histograms of prey detection times (grey circles) for a
ballistic model predator ($\mu$=1.01) with $r$=0.0003 foraging in LD
environments ($N$=5000) of varying fractal dimension at lower scales.
Foraging is destructive in all cases. Bird Island data: red triangles,
Crozet Islands: blue triangles. Recall that $D_F=\beta-1$.
(a)  $\beta$=1.2 ($\delta=5
\times 10^{-7}$, $v$=0.55), p-value of K-S test on Bird Island: 
$p_B$=0.0045, Crozet Islands: $p_C$=4.9e-08; 
(b) $\beta$=1.5 ($\delta=2  \times
10^{-4}$, $v$=0.50), $p_B$= 0.997, $p_C$=0.248; (c)
$\beta$=1.8 ($\delta=1.67 \times 10^{-3}$, $v$=0.25), $p_B$= 0.997,
$p_C$= 0.367 and (d) $\beta$=2.2 ($\delta=3.84 \times 10^{-3}$,
$v$=0.20), $p_B$= 0.033, $p_C$= 0.033.
(e)-(h): Same quantities for LD media with fixed $\beta$=1.5
($N$=5000, $\delta=2 \times 10^{-4}$) and a model forager following
processes with different step length distributions: (e) $\mu$=1.5 ($v$=0.5),
$p_B$= 1, $p_C$= 0.248; (f) $\mu$=2.0 ($v$=0.67), $p_B$= 0.999,
$p_C$= 0.0995; (g) $\mu$=2.5 ($v$=0.67), $p_B$= 0.000955, 
$p_C$=4.03e-09 and (h) $\mu$=3.0
($v$=0.67), $p_B$= 6.38e-05, $p_C$= 1.72e-13.}
\label{ld_accumulated}
\end{figure}

A ballistic walker  foraging through a LD with  $0.5\leq D_F \leq 0.9$
(corresponding to $1.5\leq \beta\leq 1.9$)
produces  a  flight duration  distribution  that  fits  very well  the
 Bird Island \cite{edwards2007revisiting} and Crozet Islands 
\cite{weimerskirch2005prey}
albatross data over  the entire range
(Figure~\ref{ld_accumulated}a-d).   Somewhat  surprisingly,   no  fine
tuning of the fractal dimension is  needed, as a range of small values
of $D_F$ describes the data equally well. In contrast, L\'evy dust landscapes
with $D_F <0.5$  or $D_F>0.9$  do not  produce  a good  agreement with  
empirical
data.  In a  given landscape,  detection patterns  are also  robust to
changes  in  the  assumptions  regarding predator  movements.  If  predators,
instead of being ballistic ($\mu \approx 1$), choose step lengths with
$1\le\mu\le 2$, for instance $1.5$ and $2$,  $p(L)$ in  
Figure~\ref{ld_accumulated}e-h
still fits the  data very well (LD with  $D_F=0.5$). Poor agreement is
obtained for  $\mu=2.5$ and larger,  therefore, albatross data  cannot
be explained by a Gaussian random walker detecting prey in a
fractal  media.  While  a   LD  fractal  prey  field  gives  detection
statistics  that are  qualitatively  in excellent  agreement with  the
observed albatross data, estimations  of $D_F$ for oceanic prey fields
are indeed  in the range  $1.2 < D_F <  1.6$ 
\cite{osborne1989fractal,tsuda1995fractal,bertrand2005levy}.  
This quantitative  difference prompts us  to analyze the FLD  model 
where good agreement can be obtained with $D_F$ in this range of values.

\begin{figure} [!ht]
\centering
\includegraphics[width=6in]{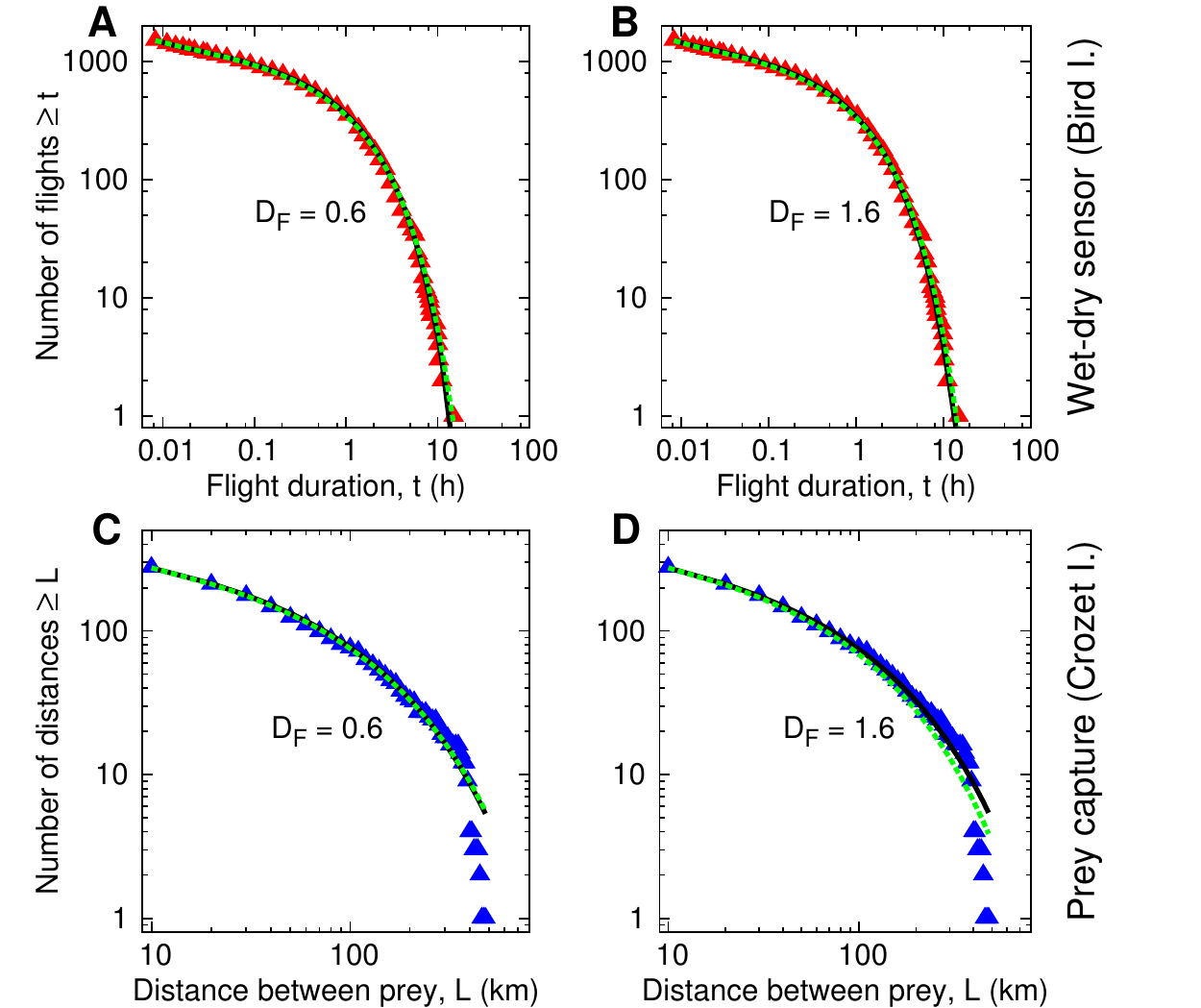}
\caption{
Cumulative distribution of prey detection times/distances
obtained by fitting the 
FLD model to the albatross data (triangles), for two fixed fractal dimensions 
of the medium ($D_F$=0.6 and 1.6). Solid black line: responsive search; green dotted line: non-responsive search. Each curve 
is plotted with the MLE of the parameters, see Table 1. 
(a)-(b): Bird Island. (c)-(d): Crozet Islands. The best estimates of
the patch size distribution parameters vary little in the different cases: 
$\nu$=1.20 $\pm$ 0.05 and $R_m$ in the range of 160-240 km, independently 
of $D_F$, for the whole range considered. 
A more efficient strategy yields a lower dimensionless detection 
radius $\tilde{r}$.
}
\label{fld_comulative} 
\end{figure}

The        FLD        model        gives        similar        results
(Figure~\ref{fld_comulative}). First, a range of values
of the fractal dimension can fit the data. Secondly, the different foraging 
behaviors considered can fit the data, too. Table 1 displays, 
for various values of
$D_F$ and  forager behaviors, the maximum  likelihood estimates (MLE)
of  the exponent $\nu$  of the  patch size distribution, of  the 
 cut-off length $R_m$, and  
of the dimensionless  detection radius $\tilde{r}$.
The responsive search scenario describes the data as well
as  the simple ballistic  one. The main difference between the two cases
is the value of $\tilde{r}$. The responsive case  is  much more  
efficient since the same prey detection patterns are obtained by a forager 
with detection radius $(\tilde{r})$ 2$-$20 times  smaller compared with 
a non-responsive forager in the same medium. 

Importantly, within each data set the MLE of  the patch size
distribution parameters  ($\nu$ and $R_m$) 
are  nearly independent of $D_F$ and the foraging scenario.
The parameter values found are also strikingly similar across the two 
albatross data sets.
Using  an  estimate of  albatrosses'  speed, 16  m/s
\cite{alerstam1993flight,weimerskirch2007does}, 
the Bird Island flight durations were converted into km.
 The values of $R_m$  in  Table 1 are of the same order of  magnitude 
 as the self-similarity
range    found    in    marine   landscapes    
\cite{tsuda1995fractal,fauchald2000scale,bertrand2005levy,makris2006fish}.  
Even   by
assuming that  the  Bird Island data are accurate  
for flights  longer than
$3\Delta  t$=30s \cite{edwards2007revisiting} or 480m,  these  values indicate
that the albatross prey field is fractal over nearly three 
logarithmic decades.

In summary, our fractal landscape model produces non-exponential 
detection patterns and can explain wandering albatross data with realistic
parameters. A non-trivial result is that
the shape of  the flight durations PDF is primarily determined by
the  patch size  distribution $\psi(R)$,  rather than  by  the fractal
dimension $D_F$. Similarly to the robustness observed in
the LD model, the  shape of $p(L)$ in the FLD model  is not altered by
modifications  in  the forager  movement  strategy.
 
\section*{Discussion}

The  foregoing results  show  the importance  of considering  predator
displacements and prey  detection events in unpredictable environments
as  two different  aspects of  the same  foraging  process.  
We emphasize that detection patterns alone are in general 
unlikely to inform on movement patterns and search strategies.
Detection
statistics  of long-ranging foraging animals in the ocean can be regarded  
as  depending on the  size of the regions with  uniform density, i.e. a
higher level of landscape organization,  and not on all the details of
the  prey field.   This result  resonates with  the current  view that
marine animals  can   track  meso  and   submeso-scale  seascape  features
\cite{tew2009top}.   Our study suggests  that detection  statistics in
both  uniform and  scale  invariant landscapes  depend  little on  the
hypothesized  predator  movement   rules,  therefore  forager  search
strategies cannot  be inferred  from detection patterns  only.
 
Wandering albatrosses adjust their movement to cope with overdispersed
preys    and    environmental    features    at    different    scales
\cite{weimerskirch2005prey,weimerskirch2007seabirds}.    The   
 two data sets analyzed here  can 
be consistently explained by different
foraging  models assuming that landings and prey capture are related 
to prey detection and
that prey  is fractally  distributed from about 200-400m up to
scales of 150$-$250 km. These scales are in  agreement with  observations  
of the distributions of pelagic fish, plankton  and squid in the 
ocean \cite{tsuda1995fractal,fauchald2000scale,makris2006fish}.
We  infer that albatross prey distribution can
be pictured as a random, self-similar assembly of regions with varying
sizes and  densities (FLD  model). The empirical PDF  of flight lengths 
is  well  reproduced if the size of the aforementioned regions follows
a  power-law  distribution  with exponent close to unity ($\nu \approx 1.2$,
see Table 1).  The truncation of
very long flights  ($>$200 km) is unavoidable as  the prey field tends
to be space filling beyond these  scales.  Our results  on landing/take-off
activity  are consistent with  direct prey  capture data  of wandering
albatrosses, suggesting  that both  are closely related,  although not
strictly  equivalent.  

The probability  distribution function  of the
local density of  krill, the prey of several  top marine predators, is
described  by   an  inverse  power-law,   $\rho^{-\alpha_\rho}$,  with
$\alpha_\rho \approx 1.7$ over four decades \cite{sims2008scaling}. 
It is likely that many other types of organisms, in particular large 
fish, follow a similar pattern \cite{makris2006fish}. In
the  FLD  model,  along  with  $\nu$, $\alpha_\rho$  is  an  important
exponent  characterizing  the  prey  field.  In the  two  examples  of
Figure~\ref{fld_comulative}-Table 1,  where $D_F$ is fixed  to 0.6 and
1.6, respectively, we obtain $\alpha_\rho$=1.75 and 1.53 from relation
(\ref{scalingrho}).  These  values  are  comparable to  the  empirical
exponent  1.7. These results also imply  considerable relative variations  
in albatross
prey density, at least of the order of $(R_m/R_0)^2 \approx 10^6$.
 
Large fluctuations in prey density  have been identified as a possible
cause      of       non-exponentially      distributed      detections
\cite{reynolds2008many}. The  FLD model  shows that it  is  indeed
the case,  if the local  density   fluctuations  are  structured   
in  widely  different characteristic sizes across many scales.
 As an example, in the ocean,
high  productivity  areas  are  separated  by larger  areas  of  lower
productivity \cite{weimerskirch2005prey, weimerskirch2007seabirds}.  A
simple  analytical calculation  can show  that a  forager  crossing an
heterogeneous  medium composed of  patches of  equal and  small sizes,
although with power-law distributed prey densities, has an exponential
$p(L)$  \cite{boyer2011inpreparation}.  Hence,  not only  prey density
distributions but the spatial arrangement of prey density fluctuations
seem to  be a crucial element to  obtain non-exponentially distributed
detections.  In a different context, the study of a model of ballistic
particles  propagating through  Sierpinski-like  fractals showed  that
detection  patterns  were  not exponentials  \cite{isliker2003random}.
Our modeled  landscapes differ from these  Sierpinski gaskets, though,
as  the fractals  considered here  are not  characterized by  a single
length  scale between nearest  prey, an  important assumption  made in
\cite{isliker2003random}.  As  noted earlier, no  general relation has
to be expected between the fractal dimension and detection statistics,
which also depend on the kind of fractal structure considered.

Our results also show that  random but uniform prey fields should lead
to   exponential  detection   patterns.  We   have  identified
exponential distributions of flight durations between dives
  in the thick-billed  murre, an Arctic  seabird that,
unlike  the   much  bigger  wandering  albatross,   forages  at  small
spatiotemporal  scales by  restricting its  search over  reduced areas
where prey predictability is higher \cite{hamish2009underwater}. These
observations  can  be interpreted  within  our  modeling framework:  a
forager with a high degree of site fidelity performing  a search restricted 
to areas  where prey encounter
is   high   should   not   experience   large   variations   in   prey
density. Therefore,  the detection patterns  should come closer  to an
exponential  form  than for  a  species  searching  over vast  oceanic
surfaces.

We conclude that detection  statistics,
along other behavioral traits of seabirds \cite{monaghan1996relevance},
 can give valuable  
information  on  the prey  field
spatial  distributions. Namely,  in  our examples the frequency
distribution of  detection times or distances follow  a
scaling law, $\lambda_d^{-1}f(L/\lambda_d)$, where $\lambda_d$
is a typical length between prey detections and depends
both on predator  movements and the prey field,  whereas $f(x)$ depends on
the prey  field only. The function $f(x)$ is typically an exponential
for uniform prey fields and may involve power-law terms for fractal media.
These  findings could be useful  for disentangling
the renewed  debate on how organism-environment  interactions build up
statistical     patterns    of     movement    
\cite{benhamou2007many,nathan2008movement,reynolds2009levy,bartumeus2009behavioral} 
not only in seabirds but in other animals as well.

\section*{Acknowledgments}
   
We appreciate financial support from PAPIIT-UNAM Grants IN-118306 and 
IN-107309. FB is supported by the Ram\'on y Cajal Program from the 
Spanish Ministry of Science and Innovation. We very much appreciate 
the kind comments from Robert May, Marcos da Luz, Gandhi Viswanathan, 
Og de Souza, Sebastian Abades, Theo Geisel and Gabriel Ramos-Fernandez.





\newpage
\section*{Tables} 

\begin{table}[!ht]
\caption{Maximum  likelihood estimates  of the  Fractal  Local Density
model parameters ($\nu$, $R_m$ and $\tilde{r}$) in several scenarii, 
for each albatross data set. (Note: in the Bird I. case, $R_m$ was obtained 
by converting flight durations into distances assuming $v=16$m/s
\cite{alerstam1993flight,weimerskirch2007does}.)}
\vspace{0.5cm}
\begin{tabular*}{0.90\textwidth}{c@{}cccccc}
$\   $  &   $\   $  &   $\nu\quad   $  &   $R_m$(km) &   $\tilde{r}$& $\epsilon$
&  ($P$, $G$)
\\ 
\hline 
{\bf Bird Island \cite{edwards2007revisiting} } &{\bf(wet-dry sensor data)} \\
\hline
$D_F$=0.6 &responsive  search  & 1.2  & 160 & 1.8 & -0.4  & (0.69, 43.0)\\ 
$"$ &non-responsive search     & 1.15 & 180 & 9.8 & -0.45 & (0.56, 46.0)\\  
$D_F$=1.6 & responsive search & 1.2  & 160 & 5.4 & -1.40 & (0.51, 47.3)\\ 
$"$ & non-responsive search    & 1.15 & 180 & 180  & -1.45  & (0.58, 45.6)\\ 
\hline
{\bf Crozet Islands \cite{weimerskirch2005prey}} &{\bf (prey capture data)}\\
\hline
$D_F$=0.6 &responsive  search  & 1.2  & 220 & 5.4 & -0.4  & (0.08, 62.3)\\ 
$"$ &non-responsive search     & 1.2 & 240 & 13.6 & -0.4 & (0.09, 61.8)\\  
$D_F$=1.6 & responsive search  & 1.25  & 240 & 6.0 & -1.35 & (0.07, 62.0)\\ 
$"$ & non-responsive search    & 1.2  & 210 & 160  & -1.40  & (0.04, 65.0)\\ 
\hline
\end{tabular*}
\end{table}

\newpage
\section*{Figure Legends}

\end{document}